# Time evolution of a Gaussian class of quasi-distribution functions under quadratic Hamiltonian

D Ginzburg[1] and A Mann[1,*]

*1 Department of Physics, Technion - Israel Institute of Technology, Haifa 32000, Israel*
*\*Corresponding author: dima.ginzburg@gmail.com*

A Lie algebraic method for propagation of the Wigner quasi-distribution function under quadratic Hamiltonian was presented by Zoubi and Ben-Aryeh. We show that the same method can be used in order to propagate a rather general class of quasi distribution functions, which we call "Gaussian class". This class contains as special cases the well-known Wigner, Husimi, Glauber and Kirkwood – Rihaczek quasi-distribution functions. We present some examples of the calculation of the time-evolution of those functions.
http://dx.doi.org/10.1364/AO.53.001648

## 1. Introduction

The idea behind quasi-distribution functions (QDF) (e.g. [1-3]) is to use a tool that resembles a classical distribution function in phase space and can be used to calculate expectation values of observables. In classical mechanics in phase space, expectation values are calculated as an integral

$$\langle A \rangle = \iint dq\, dp\, A(q,p) F(q,p,t), \quad (1.1)$$

where $\langle A \rangle$ is the expectation value of the observable $A$, $A(q,p)$ is the observable as a function of $q, p$ and $F(q,p,t)$ is the distribution function of the system in phase space.

In quantum mechanics, using the standard formulation, the expectation value of $\hat{A}(\hat{q},\hat{p})$ is calculated as

$$\langle \hat{A}(\hat{q},\hat{p}) \rangle = Tr\{\hat{\rho}(\hat{q},\hat{p},t)\hat{A}(\hat{q},\hat{p})\}. \quad (1.2)$$

We would like to find $A(q,p), F(q,p,t)$ such that

$$Tr\{\hat{\rho}(\hat{q},\hat{p},t)\hat{A}(\hat{q},\hat{p})\} = \iint dq\, dp\, A(q,p) F(q,p,t), \quad (1.3)$$

where $A(q,p)$ is a function representing the operator $\hat{A}(\hat{q},\hat{p})$ and $F(q,p,t)$ is the quasi-distribution function representing the state $\hat{\rho}$. Because of the non-zero commutation relation between $\hat{q}$ and $\hat{p}$, the mapping of $\hat{\rho}(\hat{q},\hat{p},t)$ to $F(q,p,t)$ and $\hat{A}(\hat{q},\hat{p})$ to $A(q,p)$ is not unique. To define $F(q,p,t)$ uniquely, we follow Cohen [2] and Lee [1]. We use $e^{iv\hat{q}-iu\hat{p}}$ as a generator for $\hat{A}(\hat{q},\hat{p})$, multiply by $f(v,u)$ to define the ordering, and write (we use units $\hbar = m = \omega = 1$, except when writing $\omega$ explicitly is more convenient)

$$Tr\{\hat{\rho}(\hat{q},\hat{p},t)e^{iv\hat{q}-iu\hat{p}}f(v,u)\} = \iint dq\, dp\, e^{ivq-iup} F^f(q,p,t). \quad (1.4)$$

(Lee [1] uses convention in which $\xi = v$ and $\eta = -u$.)
Using Fourier transform [1]

$$F^f(q,p,t) = \frac{1}{4\pi^2} \iint du\, dv\, Tr\{\hat{\rho}(\hat{q},\hat{p},t)e^{iv\hat{q}-iu\hat{p}}f(v,u)\}e^{-ivq+iup} = \frac{1}{4\pi^2} \iint du\, dv\, \int dq' \left\langle q'-\frac{u}{2}\middle|\hat{\rho}\middle|q'+\frac{u}{2}\right\rangle f(v,u)e^{iv(q'-q)}e^{iup} \quad (1.5)$$

Different choices of the function $f(v,u)$ correspond to different phase-space distribution functions.
The expression $\chi_f(u,v) = Tr\{\hat{\rho}(\hat{q},\hat{p},t)e^{iv\hat{q}-iu\hat{p}}f(v,u)\}$ is the characteristic function corresponding to the choice of

$f$, so the quasi-distribution function can also be written as

$$F^f(q,p,t) = \frac{1}{4\pi^2} \iint du\, dv\, \chi_f(u,v) e^{-ivq+iup}. \quad (1.6)$$

To calculate the average of an observable, we integrate

$$\langle A \rangle = \iint dq\, dp\, A^f(q,p) F^f(q,p,t) \quad (1.7)$$

where [1]

$$A^f(q,p,t) = \frac{2\pi}{4\pi^2} \iint du\, dv\, Tr\{\hat{A}(\hat{q},\hat{p},t) e^{iv\hat{q}-iu\hat{p}} f^{-1}(-v,-u)\} e^{-ivq+iup}. \quad (1.8)$$

Sometimes it is convenient to consider the quasi distribution function in complex $\alpha$ space representation (coherent state phase space representation) rather than in the $q, p$ phase space representation [1]. The two representations are related by

$$\alpha = \frac{1}{\sqrt{2}}(q+ip), \quad \alpha^* = \frac{1}{\sqrt{2}}(q-ip)$$
$$\hat{a} = \frac{1}{\sqrt{2}}(\hat{q}+i\hat{p}), \quad \hat{a}^\dagger = \frac{1}{\sqrt{2}}(\hat{q}-i\hat{p}) \quad (1.9)$$

Integration is now done by $d(\text{Re}\,\alpha)d(\text{Im}\,\alpha)$ and defined as $d^2\alpha$

$$d^2\alpha = d(\text{Re}\,\alpha)d(\text{Im}\,\alpha) = 1/2\, dq\, dp. \quad (1.10)$$

Distribution functions in $\alpha, \alpha^*$ and $q, p$ phase space representations are related through the normalization condition

$$\iint dq\, dp\, F^f(q,p,t) = \iint d^2\alpha\, F^f(\alpha,\alpha^*,t); \quad (1.11)$$

therefore,

$$F^f(\alpha,\alpha^*,t) = 2 F^f(q,p,t). \quad (1.12)$$

In order to write $F^f(\alpha,\alpha^*,t)$ in $\alpha, \alpha^*$ terms, we define new integration variables (this notation is similar to the notation of references [3-5]; Lee [1] uses $z$ instead of $\beta$)

$$\beta = \frac{1}{\sqrt{2}}(u+iv), \beta^* = \frac{1}{\sqrt{2}}(u-iv). \quad (1.13)$$

Hence,

$$F^f(\alpha,\alpha^*,t) = \frac{1}{\pi^2} \iint d^2\beta\, Tr\{\hat{\rho}(\hat{a},\hat{a}^\dagger,t) e^{\beta\hat{a}^\dagger - \beta^*\hat{a}} f(\beta,\beta^*)\} e^{-\beta\alpha^* + \beta^*\alpha}. \quad (1.14)$$

Consequently, if we define the characteristic function in coherent phase space as

$$\chi_f(\beta,\beta^*) = Tr\{\hat{\rho}(\hat{a},\hat{a}^\dagger,t) e^{\beta\hat{a}^\dagger - \beta^*\hat{a}} f(\beta,\beta^*)\}, \quad (1.15)$$

we get

$$F^f(\alpha,\alpha^*,t) = \frac{1}{\pi^2} \iint d^2\beta\, \chi_f(\beta,\beta^*) e^{-\beta\alpha^* + \beta^*\alpha}. \quad (1.16)$$

It is easy to see that

$$\chi_f = f \chi_W, \quad (1.17)$$

where $\chi_W$ is the characteristic function corresponding to symmetric ordering (Wigner function).

## 2. Dynamics

The equation of motion of a general quantum quasi-distribution function for a Hermitian Hamiltonian was first given without proof by Cohen [2]. The proof can be found in reference [1].

$$\frac{\partial F^f(q,p,t)}{\partial t} = 
2 f^{-1}\left(-i\frac{\partial}{\partial q_2}, i\frac{\partial}{\partial p_2}\right) f^{-1}\left(-i\frac{\partial}{\partial q_1}, i\frac{\partial}{\partial p_1}\right) \times 
$$
$$f\left(-i\frac{\partial}{\partial q_2} - i\frac{\partial}{\partial q_1}, i\frac{\partial}{\partial p_2} + i\frac{\partial}{\partial p_1}\right) \times$$
$$\sin\left(\frac{1}{2}\left(\frac{\partial}{\partial q_1}\frac{\partial}{\partial p_2} - \frac{\partial}{\partial q_2}\frac{\partial}{\partial p_1}\right)\right) \times$$
$$\tilde{H}^f(q_1,p_1) F^f(q_2,p_2,t) \Big|_{\substack{q_1=q_2=q,\\ p_1=p_2=p}} \quad (2.1)$$

where $\tilde{H}^f(q,p) = f\left(i\frac{\partial}{\partial q}, -i\frac{\partial}{\partial p}\right) H(q,p)$ and $H(q,p)$ is found by putting $\hat{H}(\hat{q},\hat{p})$ in symmetrical ordering and changing $\hat{q}$ to $q$ and $\hat{p}$ to $p$.

Sometimes we are interested in adding a loss mechanism to the equations. One of the ways to do it is by weak coupling to a reservoir. In this case, for the density matrix, we have the well-known master equation [6]

$$\frac{\partial \hat{\rho}(t)}{\partial t} = -i\left[\hat{H}, \hat{\rho}\right] +$$
$$\frac{\gamma}{2}(N+1)\left(2\hat{a}\hat{\rho}\hat{a}^\dagger - \hat{a}^\dagger\hat{a}\hat{\rho} - \hat{\rho}\hat{a}^\dagger\hat{a}\right) +$$
$$\frac{\gamma}{2}N\left(2\hat{a}^\dagger\hat{\rho}\hat{a} - \hat{a}\hat{a}^\dagger\hat{\rho} - \hat{\rho}\hat{a}\hat{a}^\dagger\right) + \quad . \quad (2.2)$$
$$\frac{\gamma}{2}M\left(2\hat{a}^\dagger\hat{\rho}\hat{a}^\dagger - \hat{a}^\dagger\hat{a}^\dagger\hat{\rho} - \hat{\rho}\hat{a}^\dagger\hat{a}^\dagger\right) +$$
$$\frac{\gamma}{2}M^*\left(2\hat{a}\hat{\rho}\hat{a} - \hat{a}\hat{a}\hat{\rho} - \hat{\rho}\hat{a}\hat{a}\right)$$

The operator master equation can be converted into a c-number equation for any quasi-distribution function.

We assume a quadratic Hamiltonian, i.e. Hamiltonian of the form

$$\hat{H} = \omega\left(\hat{a}^\dagger\hat{a} + 1/2\right) + \left(V\hat{a}^\dagger + V^*\hat{a}\right) + \left(A\hat{a}^\dagger\hat{a}^\dagger + A^*\hat{a}\hat{a}\right). \quad (2.3)$$

It was shown [5,7] that the equation of motion for the Wigner characteristic function is

$$\frac{\partial \chi_W(\beta, \beta^*)}{\partial t} = \begin{pmatrix} \frac{\gamma}{2}\left(-\beta\frac{\partial}{\partial\beta} - \beta^*\frac{\partial}{\partial\beta^*} - |\beta|^2\right) - \\ \gamma N |\beta|^2 - \\ \frac{\gamma}{2}\left(M\beta^{*2} + M^*\beta^2\right) - \\ i\omega\left(\beta^*\frac{\partial}{\partial\beta^*} - \beta\frac{\partial}{\partial\beta}\right) + \\ i\left(V\beta^* + V^*\beta\right) + \\ 2i\left(A\beta^*\frac{\partial}{\partial\beta} - A^*\beta\frac{\partial}{\partial\beta^*}\right) \end{pmatrix} \chi_W(\beta, \beta^*) \equiv \quad . \quad (2.4)$$
$$\hat{\Omega}_{\chi_W} \chi_W(\beta, \beta^*)$$

And for the Wigner function [5,7]

$$\frac{\partial W(\alpha, \alpha^*)}{\partial t} = \begin{pmatrix} \frac{\gamma}{2}\left(\frac{\partial}{\partial\alpha}\alpha + \frac{\partial}{\partial\alpha^*}\alpha^* + \frac{\partial^2}{\partial\alpha\partial\alpha^*}\right) + \\ \gamma N \frac{\partial^2}{\partial\alpha\partial\alpha^*} - \\ \frac{\gamma}{2}\left(M\frac{\partial^2}{\partial\alpha^2} + M^*\frac{\partial^2}{\partial\alpha^{*2}}\right) + \\ i\omega\left(\frac{\partial}{\partial\alpha}\alpha - \frac{\partial}{\partial\alpha^*}\alpha^*\right) + \\ i\left(V\frac{\partial}{\partial\alpha} - V^*\frac{\partial}{\partial\alpha^*}\right) + \\ 2i\left(A\frac{\partial}{\partial\alpha}\alpha^* - A^*\frac{\partial}{\partial\alpha^*}\alpha\right) \end{pmatrix} W(\alpha, \alpha^*) \equiv \quad . \quad (2.5)$$
$$\hat{\Omega}_W W(\alpha, \alpha^*)$$

In $q, p$ representation

$$\frac{\partial W(q, p)}{\partial t} = \begin{cases} \omega\left(\frac{\partial}{\partial p}q - \frac{\partial}{\partial q}p\right) + \frac{\gamma}{2}\left(\frac{\partial}{\partial p}p + \frac{\partial}{\partial q}q\right) + \\ \sqrt{2}\left(R\frac{\partial}{\partial p} - U\frac{\partial}{\partial q}\right) - \gamma L\frac{\partial^2}{\partial p\partial q} - \\ \frac{\gamma}{2}K\left(\frac{\partial^2}{\partial q^2} - \frac{\partial^2}{\partial p^2}\right) + \\ \frac{\gamma}{2}\left(N + \frac{1}{2}\right)\left(\frac{\partial^2}{\partial q^2} + \frac{\partial^2}{\partial p^2}\right) + \\ 2A_x\left(q\frac{\partial}{\partial p} + p\frac{\partial}{\partial q}\right) - 2A_y\left(q\frac{\partial}{\partial q} - p\frac{\partial}{\partial p}\right) \end{cases} W(q, p) = (2.6)$$
$$\hat{\Omega}_W W(q, p)$$

where $M = K + iL, \quad V = R + iU, \quad A = A_x + iA_y$.

In [4,5] Ben-Aryeh and Zoubi noticed that the equation of motion for the Wigner function can be written in the form

$$\frac{\partial F}{\partial t} = \left\{\sum_i a_i \hat{S}_i\right\} F, \quad (2.7)$$

where the operators $\hat{S}_i$ close a Lie algebra.

In $q, p$ representation the $\hat{S}_i$ are

$$\widehat{S}_1 = p\frac{\partial}{\partial q}, \quad \widehat{S}_2 = q\frac{\partial}{\partial p}, \quad \widehat{S}_3 = \frac{\partial}{\partial q}q - \frac{\partial}{\partial p}p$$

$$\widehat{S}_4 = \frac{\partial}{\partial q}q + \frac{\partial}{\partial p}p, \quad \widehat{S}_5 = \frac{\partial}{\partial q}, \quad \widehat{S}_6 = \frac{\partial}{\partial p} \quad . \quad (2.8)$$

$$\widehat{S}_7 = \frac{\partial^2}{\partial q^2}, \quad \widehat{S}_8 = \frac{\partial^2}{\partial p^2}, \quad \widehat{S}_9 = \frac{\partial^2}{\partial q \partial p}$$

Their non-zero commutation relations are

$$\begin{aligned}
&[\widehat{S}_1, \widehat{S}_2] = -\widehat{S}_3, \quad [\widehat{S}_1, \widehat{S}_3] = 2\widehat{S}_1, \quad [\widehat{S}_1, \widehat{S}_6] = -\widehat{S}_5, \\
&[\widehat{S}_1, \widehat{S}_8] = -2\widehat{S}_9, \quad [\widehat{S}_1, \widehat{S}_9] = -\widehat{S}_7, \quad [\widehat{S}_2, \widehat{S}_3] = -2\widehat{S}_2, \\
&[\widehat{S}_2, \widehat{S}_5] = -\widehat{S}_6, \quad [\widehat{S}_2, \widehat{S}_7] = -2\widehat{S}_9, \quad [\widehat{S}_2, \widehat{S}_9] = -\widehat{S}_8, \\
&[\widehat{S}_3, \widehat{S}_5] = -\widehat{S}_5, \quad [\widehat{S}_3, \widehat{S}_6] = \widehat{S}_6, \quad [\widehat{S}_3, \widehat{S}_7] = -2\widehat{S}_7, \\
&[\widehat{S}_3, \widehat{S}_8] = 2\widehat{S}_8, \quad [\widehat{S}_4, \widehat{S}_5] = -\widehat{S}_5, \quad [\widehat{S}_4, \widehat{S}_6] = -\widehat{S}_6, \\
&[\widehat{S}_4, \widehat{S}_7] = -2\widehat{S}_7, \quad [\widehat{S}_4, \widehat{S}_8] = -2\widehat{S}_8, \quad [\widehat{S}_4, \widehat{S}_9] = -2\widehat{S}_9
\end{aligned} \quad . (2.9)$$

In coherent representation, the $\widehat{S}_i$ operators are [5]

$$\widehat{S}_1 = \alpha^* \frac{\partial}{\partial \alpha}, \quad \widehat{S}_2 = \alpha \frac{\partial}{\partial \alpha^*}, \quad \widehat{S}_3 = \frac{\partial}{\partial \alpha}\alpha - \frac{\partial}{\partial \alpha^*}\alpha^*$$

$$\widehat{S}_4 = \frac{\partial}{\partial \alpha}\alpha + \frac{\partial}{\partial \alpha^*}\alpha^*, \quad \widehat{S}_5 = \frac{\partial}{\partial \alpha}, \quad \widehat{S}_6 = \frac{\partial}{\partial \alpha^*} \quad (2.10)$$

$$\widehat{S}_7 = \frac{\partial^2}{\partial \alpha^2}, \quad \widehat{S}_8 = \frac{\partial^2}{\partial \alpha^{*2}}, \quad \widehat{S}_9 = \frac{\partial^2}{\partial \alpha \partial \alpha^*}$$

and their non-zero commutation relations are the same as in equation 2.9.

It is easy to see that the propagation operator $U(t,0)$ defined by

$$F(q,p,t) = U(t,0)F(q,p,0) \quad (2.11)$$

obeys equation 2.7, and in addition equals unity when $t=0$, i.e.,

$$\frac{\partial \widehat{U}(t,0)}{\partial t} = \left\{\sum_i a_i \widehat{S}_i\right\} \widehat{U}(t,0), \quad \widehat{U}(0,0) = \widehat{1} \quad . (2.12)$$

Since the $\widehat{S}_i$ operators form a Lie algebra, the propagator can be written as [4,5,8]

$$\widehat{U}(t,0) = \prod_i \left(e^{C_i(t)\widehat{S}_i}\right) . \quad (2.13)$$

Of course different ordering of the $e^{C_i(t)\widehat{S}_i}$ will yield different $C_i$.

In this form the propagator is built as a product of simple propagators, which can be easily applied, as we will see later. To find the coefficients $C_i$ Zoubi and Ben Aryeh used a matrix representation for the operators $\widehat{S}_i$. Using this representation they found the equations for $C_i$

$$\begin{aligned}
\dot{C}_1 &= a_1 - 2a_3 C_1 - a_2 C_1^2 \\
\dot{C}_2 &= a_2 + 2a_2 C_1 C_2 + 2a_3 C_2 \\
\dot{C}_3 &= a_3 + a_2 C_1 \\
\dot{C}_4 &= a_4 \\
\dot{C}_5 &= \left(a_6 C_1 + a_5\right) e^{C_3 + C_4} \\
\dot{C}_6 &= \left(a_6 C_1 C_2 + a_5 C_2 + a_6\right) e^{C_4 - C_3} \\
\dot{C}_7 &= \left(a_8 C_1^2 + a_9 C_1 + a_7\right) e^{2C_4 + 2C_3} \\
\dot{C}_8 &= \begin{bmatrix} a_9 \left(C_2 + C_1 C_2^2\right) + \\ a_8 \left(1 + 2C_1 C_2 + C_1^2 C_2^2\right) + a_7 C_2^2 \end{bmatrix} e^{2C_4 - 2C_3} \\
\dot{C}_9 &= \left[a_9 \left(1 + 2C_1 C_2\right) + 2a_8 \left(C_1 + C_1^2 C_2\right) + 2a_7 C_2\right] e^{2C_4}
\end{aligned} \quad , (2.14)$$

with the initial condition $C_i(0) = 0$ for all $i$.

We show that the same procedure can be applied to a wider class of quasi-distribution functions.
We define the "general Gaussian quasi-distribution function" as a function for which $f$ has the form

$$f(\beta, \beta^*) = e^{A_1|\beta|^2 + A_2\beta^2 + A_3\beta^{*2}} \text{ or } f(v,u) = e^{C_1 u^2 + C_2 v^2 + C_3 2iuv} . (2.15)$$

It is easy to see that (e.g. [1])

1. For $A_1 = 0, A_2 = 0, A_3 = 0$ or $C_1 = 0, C_2 = 0, C_3 = 0$
$\chi_f$ is the characteristic function of Wigner function.

2. For $A_1 = \frac{1}{2}, A_2 = 0, A_3 = 0$ or $C_1 = \frac{1}{4}, C_2 = \frac{1}{4}, C_3 = 0$
$\chi_f$ is the characteristic function of the normal ordered function (or $P$ function).

3. For $A_1 = -\frac{1}{2}, A_2 = 0, A_3 = 0$ or
$C_1 = -\frac{1}{4}, C_2 = -\frac{1}{4}, C_3 = 0$ $\chi_f$ is a characteristic function of the anti-normal ordered function (or $Q$ function).

4. For $A_1 = 0, A_2 = -\frac{1}{4}, A_3 = \frac{1}{4}$ or $C_1 = 0, C_2 = 0, C_3 = -\frac{1}{4}$
$\chi_f$ is the characteristic function of anti-standard ordered function (or Kirkwood Rihaczek function).

5. For $A_1 = 0, A_2 = \frac{1}{4}, A_3 = -\frac{1}{4}$ or $C_1 = 0, C_2 = 0, C_3 = \frac{1}{4}$

$\chi_f$ is the characteristic function of standard ordered function (or anti Kirkwood Rihaczek function).

6. For $A_1 = \frac{s}{2}, A_2 = 0, A_3 = 0$ or $C_1 = \frac{s}{4}, C_2 = \frac{s}{4}, C_3 = 0$

$\chi_f$ is the characteristic function of the s-ordered function of Cahill and Glauber [9].

The relations between $A_1, A_2, A_3$ and $C_1, C_2, C_3$ are

$$\begin{aligned} C_1 &= \frac{1}{2}(A_1 + A_2 + A_3) & A_1 &= (C_1 + C_2) \\ C_2 &= \frac{1}{2}(A_1 - A_2 - A_3) \quad \text{or} \quad & A_2 &= \frac{1}{2}(C_1 - C_2 + 2C_3) \\ C_3 &= \frac{1}{2}(A_2 - A_3) & A_3 &= \frac{1}{2}(C_1 - C_2 - 2C_3) \end{aligned} \quad (2.16)$$

As already mentioned, Cohen presented a general equation of motion for quasi – distribution functions under a general Hamiltonian. Using his formula in the case of quadratic Hamiltonian and Gaussian quasi-distribution functions. the propagator can be decomposed into simple propagators just like the case of the Wigner function. (In this section we show that it is true for propagation without damping; for propagation with damping see the appendix).
Here we do the calculation only in $q, p$ representation.
The quadratic Hamiltonian is

$$\hat{H}(\hat{q}, \hat{p}) = K_1 \hat{q}^2 + K_2 \hat{p}^2 + K_3 \frac{1}{2}(\hat{q}\hat{p} + \hat{p}\hat{q}) + K_4 \hat{q} + K_5 \hat{p} \quad . \quad (2.17)$$

We use equations 2.1, 2.15 and 2.17 and obtain

$$\tilde{H}^f(q_1, p_1) = K_1 q_1^2 + K_2 p_1^2 + K_3 q_1 p_1 + K_4 q_1 + K_5 p_1 - 2C_2 K_1 - 2C_1 K_2 + 2C_3 i K_3 \quad (2.18)$$

and the equation of motion

$$\begin{aligned} \frac{\partial F^f(q, p, t)}{\partial t} =& -2K_2 p \frac{\partial F^f(q, p, t)}{\partial q} + \\ & 2K_1 q \frac{\partial F^f(q, p, t)}{\partial p} + \\ & K_3 \left( p \frac{\partial F^f(q, p, t)}{\partial p} - q \frac{\partial F^f(q, p, t)}{\partial q} \right) + \\ & (2K_3 C_2 - 4iK_2 C_3) \frac{\partial^2 F^f(q, p, t)}{\partial q^2} + \\ & (4iK_1 C_3 - 2K_3 C_1) \frac{\partial^2 F^f(q, p, t)}{\partial p^2} + \\ & (4K_2 C_1 - 4K_1 C_2) \frac{\partial^2 F^f(q, p, t)}{\partial p \partial q} + \\ & K_4 \frac{\partial F^f(q, p, t)}{\partial p} - K_5 \frac{\partial F^f(q, p, t)}{\partial q} \end{aligned} \quad . \quad (2.19)$$

We see that the equation of motion has, again, the form

$$\frac{\partial F(q, p, t)}{\partial t} = \left\{ \sum_i a_i \hat{S}_i \right\} F(q, p, t),$$

where the coefficients $a_i$ are

$$\begin{aligned} a_1 &= -2K_2, & a_2 &= 2K_1, & a_3 &= -K_3, & a_4 &= 0, \\ a_5 &= -K_5, & a_6 &= K_4, & a_7 &= 2K_3 C_2 - 4iK_2 C_3, \\ a_8 &= -2K_3 C_1 + 4iK_1 C_3, & a_9 &= -4K_1 C_2 + 4K_2 C_1 \end{aligned} \quad (2.20)$$

Therefore, the same technique which Zoubi and Ben Aryeh used to decompose the propagator of the Wigner function into simple propagators can be applied here, too. The propagation operator may be written as
$\hat{U}(t, 0) = \prod_i \left( e^{C_i(t) \hat{S}_i} \right)$, where the $C_i(t)$ are related to the

$a_i(t)$ via equations 2.14.

### 3. Examples

#### 3.1 Free particle ($\hat{H}(\hat{q}, \hat{p}) = \frac{\hat{p}^2}{2}$)

##### 3.1.1 Wigner function (free particle)
The equation of motion for the Wigner function is

$$\frac{\partial F^W(q, p, t)}{\partial t} = -p \frac{\partial}{\partial q} F^W(q, p, t). \quad (3.1)$$

In this case, the solution is immediate

$$F^W(q, p, t) = e^{-pt \frac{\partial}{\partial q}} F^W(q, p, 0) = F^W(q - pt, p, 0). \quad (3.2)$$

### 3.1.2 Standard ordered (anti – Rihaczek) function (free particle)

In this case, the equation of motion is

$$\frac{\partial F^S(q,p,t)}{\partial t} = -p\frac{\partial}{\partial q}F^S(q,p,t) - i\frac{1}{2}\frac{\partial^2 F^S(q,p,t)}{\partial q^2}. \quad (3.3)$$

Since the operators $p\frac{\partial}{\partial q}$ and $\frac{\partial^2}{\partial q^2}$ commute with each other, the solution is again immediate

$$F^S(q,p,t) = e^{-pt\frac{\partial}{\partial q} - \frac{i}{2}\frac{\partial^2}{\partial q^2}t}F^S(q,p,0) = \\ e^{-pt\frac{\partial}{\partial q}}e^{-\frac{i}{2}\frac{\partial^2}{\partial q^2}t}F^S(q,p,0). \quad (3.4)$$

Since only derivatives with respect to $q$ appear in the propagator, the propagator has a very simple form in Fourier space and so

$$F^S(q,p,t) = \frac{1}{2\pi}\int e^{i\left(\frac{1}{2}k^2 - pk\right)t + ikq - ikq'} F^S(q',p,0)\,dq'\,dk. \quad (3.5)$$

For example, for $\psi(q,0) = \pi^{-1/4}e^{-q^2/2}$, the above integral yields

$$F^S(q,p,t) = \frac{1}{\sqrt{2\pi}}\frac{1}{\sqrt{\pi}}e^{iqp}(1-it)^{-1/2}e^{-\frac{q^2}{2(1-it)}}e^{-\frac{p^2}{2}(1+it)}. \quad (3.6)$$

### 3.1.3 Anti-normal quasi-distribution function (free particle)

In this case, the equation of motion is

$$\frac{\partial F^{AN}(q,p,t)}{\partial t} = \\ -p\frac{\partial}{\partial q}F^{AN}(q,p,t) - \frac{1}{2}\frac{\partial^2 F^{AN}(q,p,t)}{\partial q \partial p}. \quad (3.7)$$

In this case the operators do not commute. We see that $a_1 = -1, a_9 = -\frac{1}{2}$. The differential equations are

$$\dot{C}_1 = -1,\ \dot{C}_7 = -\frac{1}{2}C_1,\ \dot{C}_9 = -\frac{1}{2} \quad \text{with} \quad C_i(0) = 0. \quad (3.8)$$

Their solutions are

$$C_1 = -t,\quad C_7 = \frac{1}{4}t^2,\quad C_9 = -\frac{1}{2}t. \quad (3.9)$$

So we get

$$F^{AN}(q,p,t) = e^{-tp\frac{\partial}{\partial q}}e^{\frac{1}{4}t^2\frac{\partial^2}{\partial q^2}}e^{-\frac{1}{2}t\frac{\partial^2}{\partial q \partial p}}F^{AN}(q,p,0). \quad (3.10)$$

To apply the propagator, it is convenient to write $F^{AN}(q,p,0)$ as a Fourier transform, and then

$$F^{AN}(q,p,t) = e^{-tp\frac{\partial}{\partial q}}e^{\frac{1}{4}t^2\frac{\partial^2}{\partial q^2}}e^{-\frac{1}{2}t\frac{\partial^2}{\partial q \partial p}} \times \\ \frac{1}{4\pi^2}\int e^{ipu - ivq}\chi_{AN}(u,v,0)\,dv\,du = \\ \frac{1}{4\pi^2}\int e^{-tp(-iv) + \frac{1}{4}t^2(-iv)^2 - \frac{1}{2}t(-iv)(iu) + ipu - ivq}\chi_{AN}(u,v,0)\,dv\,du \quad (3.11)$$

In order to demonstrate the use of the propagator, we calculate the anti-normal QDF of $\psi(q,0) = \pi^{-1/4}e^{-q^2/2}$ as a function of time. The anti-normal characteristic function for this state is

$$\chi_{AN}(u,v,0) = e^{-|\beta|^2} = e^{-\frac{u^2+v^2}{2}}. \quad (3.12)$$

And so the anti-normal quasi-distribution function is

$$F^{AN}(q,p,t) = \\ \frac{1}{4\pi^2}\int e^{-tp(-iv) + \frac{1}{4}t^2(-iv)^2 - \frac{1}{2}t(-iv)(iu) + ipu - ivq - \frac{v^2+u^2}{2}}\,dv\,du = \\ \frac{1}{\pi}\frac{1}{\sqrt{4+t^2}}e^{\frac{-2p^2 - 2q^2 - p^2t^2 + 2qpt}{4+t^2}} \quad (3.13)$$

### 3.2 Simple harmonic oscillator ($\hat{H}(\hat{q},\hat{p}) = \frac{\hat{q}^2}{2} + \frac{\hat{p}^2}{2}$)

### 3.2.1 Wigner function (Harmonic oscillator)

In this case, the equation of motion is

$$\frac{\partial F^W(q,p,t)}{\partial t} = -p\frac{\partial}{\partial q}F^W(q,p,t) + q\frac{\partial}{\partial p}F^W(q,p,t). \quad (3.14)$$

The only non-zero coefficients are $a_1 = -1, a_2 = 1$
So the equations for the $C_i$ are

$$\dot{C}_1 = a_1 - a_2 C_1^2 \\ \dot{C}_2 = a_2 + 2a_2 C_1 C_2. \quad (3.15) \\ \dot{C}_3 = a_2 C_1$$

The solutions for these equations are

$$C_1 = -\tan(t),$$
$$C_2 = \sin(t)\cos(t), \quad (3.16)$$
$$C_3 = \ln(\cos(t))$$

And so the Wigner function as a function of time is

$$W(q,p,t) = e^{-\tan(t)p\frac{\partial}{\partial q}} e^{\sin(t)\cos(t)q\frac{\partial}{\partial p}} e^{\ln(\cos(t))\left(\frac{\partial}{\partial q}q - \frac{\partial}{\partial p}p\right)} W(q,p,0) . \quad (3.17)$$

It is easy to prove that

$$e^{-a\frac{\partial}{\partial q}} F(q,p) = F(q-a,p),$$
$$e^{-a\frac{\partial}{\partial p}} F(q,p) = F(q,p-a), \quad (3.18)$$
$$e^{a\left(\frac{\partial}{\partial q}q - \frac{\partial}{\partial p}p\right)} F(q,p,t) = F(e^a q, e^{-a} p)$$

Hence, we get

$$W(q,p,t) = e^{-\tan(t)p\frac{\partial}{\partial q}} e^{\sin(t)\cos(t)q\frac{\partial}{\partial p}} e^{\ln(\cos(t))\left(\frac{\partial}{\partial q}q - \frac{\partial}{\partial p}p\right)} W(q,p,0) = \quad (3.19)$$
$$W\begin{pmatrix}\cos(t)(q-\tan(t)p), \\ \frac{1}{\cos(t)}(p+\sin(t)\cos(t)(q-\tan(t)p)), 0\end{pmatrix}$$

In order to see the behavior at $\omega \to 0$, we write the previous equation in standard units:

$$W(q,p,t) = $$
$$W\begin{pmatrix}\cos(\omega t)\left(q-\frac{\tan(\omega t)}{m\omega}p\right), \\ \frac{\left(p+m\omega\sin(\omega t)\cos(\omega t)\left(q-\frac{\tan(\omega t)}{m\omega}p\right)\right)}{\cos(\omega t)}, 0\end{pmatrix} . \quad (3.20)$$

It is easy to see that for $\omega \to 0$, $W(q,p,t)$ tends to $W\left(q-\frac{p}{m}t, p, 0\right)$ as it should.

### 3.2.2 Standard quasi-distribution function (Harmonic oscillator)

The equation of motion is

$$\frac{\partial F^S(q,p,t)}{\partial t} = -p\frac{\partial}{\partial q}F^S(q,p,t) + q\frac{\partial}{\partial p}F^S(q,p,t) - \frac{i}{2}\frac{\partial^2}{\partial q^2}F^S(q,p,t) + \frac{i}{2}\frac{\partial^2}{\partial p^2}F^S(q,p,t) \quad (3.21)$$

The $a_i$ coefficients are

$$a_1 = -1, \quad a_2 = 1, \quad a_7 = -\frac{i}{2}, \quad a_8 = \frac{i}{2}. \quad (3.22)$$

Solving these equations we get

$$C_1 = -\tan(t), \quad C_2 = \sin(t)\cos(t),$$
$$C_3 = \ln(\cos(t)), \quad C_7 = -\frac{i}{4}\sin(2t), \quad (3.23)$$
$$C_8 = \frac{i}{4}\sin(2t), \quad C_9 = \frac{i}{2}(\cos(2t)-1)$$

Thus, the standard quasi-distribution function as a function of time is

$$F^S(q,p,t) = e^{-\tan(t)p\frac{\partial}{\partial q}} e^{\sin(t)\cos(t)q\frac{\partial}{\partial p}} e^{\ln(\cos(t))\left(\frac{\partial}{\partial q}q - \frac{\partial}{\partial p}p\right)} \times$$
$$e^{-\frac{i}{4}\sin(2t)\frac{\partial^2}{\partial q^2}} e^{\frac{i}{4}\sin(2t)\frac{\partial^2}{\partial p^2}} e^{\frac{i}{2}(\cos(2t)-1)\frac{\partial^2}{\partial q\partial p}} F^S(q,p,0) \quad (3.24)$$

As an example we show the standard QDF as a function of time for some states

#### 3.2.2.1 Superposition of the ground state and the first excited state

The standard quasi-distribution function for a superposition of the ground state and the first excited state at $t=0$ is

$$F^S_{\frac{1}{\sqrt{2}}|0\rangle+\frac{1}{\sqrt{2}}|1\rangle}(q,p,0) =$$
$$\frac{1}{\sqrt{2\pi}} e^{iqp}\left(\frac{1}{\sqrt{2}}\pi^{-1/4} e^{-q^2/2} + \frac{1}{\sqrt{2}}\pi^{-1/4}\sqrt{2}q e^{-q^2/2}\right) \times$$
$$\left(\frac{1}{\sqrt{2}}\pi^{-1/4} e^{-p^2/2} - i\frac{1}{\sqrt{2}}\pi^{-1/4}\sqrt{2}p e^{-p^2/2}\right) = \quad (3.25)$$
$$\frac{1}{2}\frac{1}{\sqrt{2}}\frac{1}{\pi} e^{iqp} e^{-q^2/2} e^{-p^2/2} - i\frac{1}{2}\frac{1}{\pi} e^{iqp} e^{-q^2/2} p e^{-p^2/2} +$$
$$\frac{1}{2}\frac{1}{\pi} q e^{-q^2/2} e^{-p^2/2} e^{iqp} - i\frac{1}{\sqrt{2}}\frac{1}{\pi} q e^{-q^2/2} p e^{-p^2/2} e^{iqp}$$

Propagating the function gives

$$F^S_{\frac{1}{\sqrt{2}}|0\rangle+\frac{1}{\sqrt{2}}|1\rangle}(q,p,t) =$$

$$\frac{1}{\sqrt{2\pi}} e^{iqp} \left( \begin{array}{c} \frac{1}{\sqrt{2}} \pi^{-1/4} e^{-q^2/2} e^{i\frac{1}{2}t} + \\ \frac{1}{\sqrt{2}} \pi^{-1/4} \sqrt{2} q e^{-q^2/2} e^{i\frac{3}{2}t} \end{array} \right) \times$$

$$\times \left( \frac{1}{\sqrt{2}} \pi^{-1/4} e^{-p^2/2} e^{-i\frac{1}{2}t} - i \frac{1}{\sqrt{2}} \pi^{-1/4} \sqrt{2} p e^{-p^2/2} e^{-i\frac{3}{2}t} \right) =$$

$$\frac{1}{\sqrt{2}} \frac{1}{2\pi} e^{iqp} e^{-q^2/2} e^{-p^2/2} - i \frac{1}{2\pi} e^{iqp} e^{-q^2/2} p e^{-p^2/2} e^{-it} + \quad (3.26)$$

$$\frac{1}{2\pi} q e^{iqp} e^{-q^2/2} e^{-p^2/2} e^{it} - i \frac{1}{\sqrt{2}} \frac{1}{\pi} e^{iqp} q e^{-q^2/2} p e^{-p^2/2} =$$

$$\frac{1}{2} \frac{1}{\sqrt{2}} \frac{1}{\pi} e^{iqp} e^{-q^2/2} e^{-p^2/2} - i \frac{1}{\sqrt{2}} \frac{1}{\pi} e^{iqp} q e^{-q^2/2} p e^{-p^2/2} +$$

$$\frac{1}{2\pi} e^{iqp} e^{-q^2/2} p e^{-p^2/2} \left( \begin{array}{c} q\cos(t) + iq\sin(t) - \\ ip\cos(t) - p\sin(t) \end{array} \right)$$

### 3.2.2.2 Cat state

A cat state is defined as

$$|\psi_{cat}\rangle = \frac{1}{\sqrt{2(1+e^{-2|\alpha|^2})}} (|\alpha\rangle + |-\alpha\rangle). \quad (3.27)$$

The standard ordered quasi-distribution function of the cat state at $t=0$ is

$$F^S_{cat}(q,p,0) = \frac{1}{\sqrt{2\pi}} e^{iqp} \psi^*(q) \tilde{\psi}(p) =$$

$$\frac{1}{\sqrt{2\pi}} e^{iqp} \frac{1}{2+2e^{-|\alpha|^2}} \left( \begin{array}{c} (\langle\alpha|q\rangle + \langle-\alpha|q\rangle) \times \\ (\langle p|\alpha\rangle + \langle p|-\alpha\rangle) \end{array} \right). \quad (3.28)$$

Using

$$\langle q|\alpha\rangle = \left(\frac{1}{\pi}\right)^{\frac{1}{4}} e^{-\frac{1}{2}(q-\sqrt{2}A)^2 + \sqrt{2}qiB - ABi},$$

$$\langle p|\alpha\rangle = \left(\frac{1}{\pi}\right)^{\frac{1}{4}} e^{-\frac{1}{2}(p-\sqrt{2}B)^2 + iAB - \sqrt{2}Api} \quad (3.29)$$

where $\alpha = A+iB$, we obtain

$$F^S_{cat}(q,p,0) = \frac{1}{\sqrt{2\pi}} \frac{1}{2+2e^{-2|\alpha|^2}} \frac{1}{\sqrt{\pi}} e^{2iAB} \times$$

$$\left( \begin{array}{c} e^{iqp} e^{-\frac{1}{2}(q-\sqrt{2}A)^2 - \frac{1}{2}(p-\sqrt{2}B)^2 - \sqrt{2}i(qB+Ap)} + \\ e^{iqp} e^{-\frac{1}{2}(q-\sqrt{2}A)^2 - \frac{1}{2}(p+\sqrt{2}B)^2 - \sqrt{2}i(qB-Ap)} + \\ e^{iqp} e^{-\frac{1}{2}(q+\sqrt{2}A)^2 - \frac{1}{2}(p-\sqrt{2}B)^2 - \sqrt{2}i(-qB+Ap)} + \\ e^{iqp} e^{-\frac{1}{2}(q+\sqrt{2}A)^2 - \frac{1}{2}(p+\sqrt{2}B)^2 + \sqrt{2}i(qB+Ap)} \end{array} \right). \quad (3.30)$$

Performing the propagation, we get

$$F^S_{cat}(q,p,t) = \frac{1}{\sqrt{2\pi}} \frac{1}{2+2e^{-2|\alpha|^2}} e^{2iAB(2\cos^2 t - 1)} \times$$

$$e^{i(B^2-A^2)\sin 2t} e^{iqp} \left( \begin{array}{c} e^{-\frac{1}{2}(q-\sqrt{2}N)^2 - \frac{1}{2}(p-\sqrt{2}M)^2 - \sqrt{2}i(qM+Np)} + \\ e^{-\frac{1}{2}(q-\sqrt{2}N)^2 - \frac{1}{2}(p+\sqrt{2}M)^2 - \sqrt{2}i(qM-Np)} + \\ e^{-\frac{1}{2}(q+\sqrt{2}N)^2 - \frac{1}{2}(p-\sqrt{2}M)^2 - \sqrt{2}i(-qM+Np)} + \\ e^{-\frac{1}{2}(q+\sqrt{2}N)^2 - \frac{1}{2}(p+\sqrt{2}M)^2 + \sqrt{2}i(qM+Np)} \end{array} \right) \quad (3.31)$$

where $N = A\cos t + B\sin t$ and $M = B\cos t - A\sin t$.

### 3.2.3 Normal and anti-normal distribution function (Harmonic oscillator)

Using equations 2.19, 2.20 we see that if $K_1 = K_2$ (i.e. the coefficients of $\hat{p}^2$ and $\hat{q}^2$ are the same) we get very simple equations with only two operators for the normal and anti – normal distribution functions, which are the same as the equation of the Wigner function (see equation 3.14)

$$\frac{\partial F^{N/AN}(q,p,t)}{\partial t} = -p \frac{\partial F^{N/AN}(q,p,t)}{\partial q} + q \frac{\partial F^{N/AN}(q,p,t)}{\partial p}. \quad (3.32)$$

This can be always achieved by proper change of variables, but just in case we have only one oscillator frequency in the system. For others the equation will also contain the term $\frac{\partial^2 F^{N/AN}(q,p,t)}{\partial p \partial q}$. Consequently the equation of motion for the normal and anti – normal distribution functions in this case of $K_1 = K_2$ will be as in equation of 3.19, which is

$$F^{N/AN}(q,p,t) =$$

$$e^{-\tan(t)p\frac{\partial}{\partial q}} e^{\sin(t)\cos(t)q\frac{\partial}{\partial p}} e^{\ln(\cos(t))\left(\frac{\partial}{\partial q}q - \frac{\partial}{\partial p}p\right)} F^{N/AN}(q,p,0). \quad (3.33)$$

Hence, similarly to the Wigner function, we obtain

$$F(q,p,t) = F\begin{pmatrix} \cos(t)(q-\tan(t)p), \\ \dfrac{1}{\cos(t)}(p+\sin(t)\cos(t)(q-\tan(t)p)),0 \end{pmatrix}. \quad (3.34)$$

### 3.3 The Hamiltonian $\hat{H} = \varepsilon e^{-2\delta t}\hat{q}^2 + \varepsilon e^{2\delta t}\hat{p}^2 + \dfrac{\delta}{2}(\hat{q}\hat{p}+\hat{p}\hat{q})$

#### 3.3.1 Standard quasi-distribution function

In this case we look at an example of a Hamiltonian with time dependent coefficients. The above Hamiltonian represents a simple harmonic oscillator with a mass, which exponentially changes in time and with a squeezing term. We use equations 2.19, 2.20, and get the equation of motion for the standard quasi-distribution function for the above Hamiltonian

$$\frac{\partial F^S(q,p,t)}{\partial t} =$$
$$-2\varepsilon e^{2\delta t} p \frac{\partial}{\partial q} F^S(q,p,t) + 2\varepsilon e^{-2\delta t} q \frac{\partial}{\partial p} F^S(q,p,t) -$$
$$i\varepsilon e^{2\delta t} \frac{\partial^2}{\partial q^2} F^S(q,p,t) + i\varepsilon e^{-2\delta t} \frac{\partial^2}{\partial p^2} F^S(q,p,t) +$$
$$\delta\left(p\frac{\partial}{\partial p}F^S(q,p,t) - q\frac{\partial}{\partial q}F^S(q,p,t)\right) \quad (3.35)$$

The non-zero coefficients $a_i$ are

$$a_1 = -2\varepsilon e^{2\delta t}, \quad a_2 = 2\varepsilon e^{-2\delta t}, \quad a_7 = -i\varepsilon e^{2\delta t},$$
$$a_8 = i\varepsilon e^{-2\delta t}, \quad a_3 = -\delta. \quad (3.36)$$

The differential equations for the $C_i$ coefficients are

$$\dot{C}_1 = a_1 - 2a_3 C_1 - a_2 C_1^2$$
$$\dot{C}_2 = a_2 + 2a_2 C_1 C_2 + 2a_3 C_2$$
$$\dot{C}_3 = a_3 + a_2 C_1$$
$$\dot{C}_7 = (a_8 C_1^2 + a_7)e^{2C_3} \quad . \quad (3.37)$$
$$\dot{C}_8 = (a_8(1+2C_1C_2+C_1^2C_2^2) + a_7 C_2^2)e^{-2C_3}$$
$$\dot{C}_9 = (2a_8(C_1+C_1^2C_2) + 2a_7 C_2)$$

And their solutions are

$$C_1 = -\tan(2\varepsilon t)e^{2\delta t}, \quad C_2 = e^{-2\delta t}\sin(2\varepsilon t)\cos(2\varepsilon t),$$
$$C_3 = -\delta t + \ln(\cos(2\varepsilon t)), \quad C_7 = \frac{-i\sin(4\varepsilon t)}{4}, \quad .(3.38)$$
$$C_8 = \frac{i}{4}\sin(4\varepsilon t), \quad C_9 = \frac{i}{2}(\cos(4\varepsilon t)-1)$$

Hence the standard ordered quasi-distribution function as a function of time is

$$F^S(q,p,t) =$$
$$e^{-\tan(2\varepsilon t)e^{2\delta t}p\frac{\partial}{\partial q}} e^{e^{-2\delta t}\sin(2\varepsilon t)\cos(2\varepsilon t)q\frac{\partial}{\partial p}} e^{(-\delta t + \ln(\cos(2\varepsilon t)))\left(\frac{\partial}{\partial q}q - \frac{\partial}{\partial p}p\right)} \times .(3.39)$$
$$e^{\frac{-i\sin(4\varepsilon t)}{4}\frac{\partial^2}{\partial q^2}} e^{\frac{i}{4}\sin(4\varepsilon t)\frac{\partial^2}{\partial p^2}} e^{\frac{i}{2}(\cos(4\varepsilon t)-1)\frac{\partial^2}{\partial q\partial p}} F^S(q,p,0)$$

For example, if we propagate the standard QDF of a harmonic ground state for the above Hamiltonian, we get

$$F^S(q,p,t) = \frac{1}{\sqrt{2\pi}}\frac{1}{\sqrt{\pi}}e^{ipq-\frac{q^2}{2}e^{-2\delta t}-\frac{p^2}{2}e^{2\delta t}}. \quad (3.40)$$

### 4. Summary and conclusions

We have presented a Lie algebraic approach to the time propagation of the general Gaussian quasi-distribution function evolving under a general quadratic Hamiltonian (and including damping). It seems that the underlying reason for the possibility to use this approach may be traced to
1. The fact that the operators involved in the Hamiltonian (equation 2.19) and in the damping (equations A.8, A.11) close the general Harmonic Oscillator Lie algebra.
2. The choice of the function $f$ (equation 2.15).

We expect that this approach may be generalized to treat the time-development of classes of quasi-distribution functions evolving under Hamiltonians which involve a finite Lie algebra.

### Appendix

Our purpose is to show that adding damping to a general Gaussian quasi-distribution function, will still yield just the same operators that appeared in the case of the Wigner function. Since the damping is more easily written in the coherent $\alpha$ representation, we will use it here. We find the changes in equations 2.4 - 2.6 when using a general Gaussian quasi-distribution.
In coherent space, the Wigner function is defined as

$$W(\alpha,\alpha^*) = \frac{1}{\pi^2}\int e^{\alpha\beta^* - \alpha^*\beta}\chi_W(\beta,\beta^*)d^2\beta. \quad (A.1)$$

Its evolution under quadratic Hamiltonian with additional damping is

$$\frac{\partial}{\partial t}W(\alpha,\alpha^*) = \frac{1}{\pi^2}\int e^{\alpha\beta^*-\alpha^*\beta}\frac{\partial \chi_W(\beta,\beta^*)}{\partial t}d^2\beta = \frac{1}{\pi^2}\int e^{\alpha\beta^*-\alpha^*\beta}\hat{\Omega}_{\chi_W}\chi_W(\beta,\beta^*)d^2\beta \quad (A.2)$$

Since the definition of a general quasi-distribution function is

$$F^f(\alpha,\alpha^*) = \frac{1}{\pi^2}\int e^{\alpha\beta^*-\alpha^*\beta}f(\beta,\beta^*)\chi_W(\beta,\beta^*)d^2\beta, \quad (A.3)$$

its evolution in time under a quadratic Hamiltonian with additional damping is

$$\begin{aligned}\frac{\partial F^f(\alpha,\alpha^*)}{\partial t} &= \\ \frac{1}{\pi^2}\int e^{\alpha\beta^*-\alpha^*\beta}f(\beta,\beta^*)\frac{\partial \chi_W(\beta,\beta^*)}{\partial t}d^2\beta &= \\ \frac{1}{\pi^2}\int e^{\alpha\beta^*-\alpha^*\beta}f(\beta,\beta^*)\hat{\Omega}_{\chi_W}\chi_W(\beta,\beta^*)d^2\beta &= \\ \frac{\int e^{\alpha\beta^*-\alpha^*\beta}f(\beta,\beta^*)\hat{\Omega}_{\chi_W}\left(\chi_f(\beta,\beta^*)f(\beta,\beta^*)^{-1}\right)d^2\beta}{\pi^2}&\end{aligned} \quad (A.4)$$

Substituting $\hat{\Omega}_{\chi_W}$ we get

$$\frac{\partial F^f(\alpha,\alpha^*)}{\partial t} = \frac{1}{\pi^2}\int e^{\alpha\beta^*-\alpha^*\beta}\hat{\Omega}_{\chi_W}\left(\chi_f(\beta,\beta^*)\right)d^2\beta + \frac{1}{\pi^2}\int e^{\alpha\beta^*-\alpha^*\beta}\chi_f(\beta,\beta^*)f(\beta,\beta^*)\hat{\Gamma}f(\beta,\beta^*)^{-1}d^2\beta, \quad (A.5)$$

where

$$\hat{\Gamma} = \begin{pmatrix} -i\omega\left(\beta^*\frac{\partial}{\partial \beta^*} - \beta\frac{\partial}{\partial \beta}\right) + \frac{\gamma}{2}\left(-\beta\frac{\partial}{\partial \beta} - \beta^*\frac{\partial}{\partial \beta^*}\right) + \\ 2i\left(A\beta^*\frac{\partial}{\partial \beta} - A^*\beta\frac{\partial}{\partial \beta^*}\right) \end{pmatrix}.$$

It is easy to show that

$$\frac{1}{\pi^2}\int e^{\alpha\beta^*-\alpha^*\beta}\hat{\Omega}_{\chi_W}\left(\chi_f(\beta,\beta^*)\right)d^2\beta = \hat{\Omega}_W F^f(\alpha,\alpha^*). \quad (A.6)$$

We substitute the expression for a Gaussian $f(\beta,\beta^*)$ (equation 2.15) and find for the second integral in A.5

$$\begin{aligned} f(\beta,\beta^*)\left(-i\omega\left(\beta^*\frac{\partial}{\partial \beta^*} - \beta\frac{\partial}{\partial \beta}\right)\right)f(\beta,\beta^*)^{-1} &= \\ 2i\omega\left(A_3\beta^{*2} - A_2\beta^2\right),&\\ f(\beta,\beta^*)\left(\frac{\gamma}{2}\left(-\beta\frac{\partial}{\partial \beta} - \beta^*\frac{\partial}{\partial \beta^*}\right)\right)f(\beta,\beta^*)^{-1} &= \\ \gamma\left(A_1|\beta|^2 + A_2\beta^2 + A_3\beta^{*2}\right),&\\ f(\beta,\beta^*)\left(2i\left(A\beta^*\frac{\partial}{\partial \beta} - A^*\beta\frac{\partial}{\partial \beta^*}\right)\right)f(\beta,\beta^*)^{-1} &= \\ 2i\left(A^*\left(A_1\beta^2 + 2A_3|\beta|^2\right) - A\left(A_1\beta^{*2} + 2A_2|\beta|^2\right)\right)&\end{aligned} \quad (A.7)$$

Hence,

$$\begin{aligned}\frac{\partial F^f(\alpha,\alpha^*)}{\partial t} &= \hat{\Omega}_W F^f(\alpha,\alpha^*) + \\ &\int \frac{e^{\alpha\beta^*-\alpha^*\beta}}{\pi^2}\chi_f(\beta,\beta^*)2i\omega\left(A_3\beta^{*2} - A_2\beta^2\right)d^2\beta + \\ &\int \frac{e^{\alpha\beta^*-\alpha^*\beta}}{\pi^2}\chi_f(\beta,\beta^*)\gamma\left(A_1|\beta|^2 + A_2\beta^2 + A_3\beta^{*2}\right)d^2\beta + \\ &\int \frac{e^{\alpha\beta^*-\alpha^*\beta}}{\pi^2}\chi_f(\beta,\beta^*)2i\begin{pmatrix}A^*\left(A_1\beta^2 + 2A_3|\beta|^2\right) - \\ A\left(A_1\beta^{*2} + 2A_2|\beta|^2\right)\end{pmatrix}d^2\beta = \\ &\hat{\Omega}_W F^f(\alpha,\alpha^*) + \\ &\begin{pmatrix} 2i\omega A_3\frac{\partial^2}{\partial \alpha^2} - 2i\omega A_2\frac{\partial^2}{\partial \alpha^{*2}} + 2iA^*A_1\frac{\partial^2}{\partial \alpha^{*2}} - \\ 4iA^*A_3\frac{\partial^2}{\partial \alpha\partial \alpha^*} - 2iAA_1\frac{\partial^2}{\partial \alpha^2} + 4iAA_2\frac{\partial^2}{\partial \alpha\partial \alpha^*} \\ +\gamma\left(-A_1\frac{\partial^2}{\partial \alpha\partial \alpha^*} + A_2\frac{\partial^2}{\partial \alpha^{*2}} + A_3\frac{\partial^2}{\partial \alpha^2}\right) \end{pmatrix} F^f(\alpha,\alpha^*)\end{aligned} \quad (A.8)$$

In the last identity we used the definition

$$F^f(\alpha,\alpha^*) = \frac{1}{\pi^2}\int e^{\alpha\beta^*-\alpha^*\beta}\chi_f(\beta,\beta^*)d^2\beta \quad (A.9)$$

and the identities

$$\begin{aligned}-\frac{\partial}{\partial \alpha^*}\frac{\partial}{\partial \alpha}F(\alpha,\alpha^*) &= \frac{1}{\pi^2}\int e^{\alpha\beta^*-\alpha^*\beta}\beta\beta^*\chi_f(\beta,\beta^*)d^2\beta \\ \frac{\partial^2}{\partial \alpha^{*2}}F(\alpha,\alpha^*) &= \frac{1}{\pi^2}\int e^{\alpha\beta^*-\alpha^*\beta}\beta^2\chi_f(\beta,\beta^*)d^2\beta \\ \frac{\partial^2}{\partial \alpha^2}F(\alpha,\alpha^*) &= \frac{1}{\pi^2}\int e^{\alpha\beta^*-\alpha^*\beta}\beta^{*2}\chi_f(\beta,\beta^*)d^2\beta\end{aligned} \quad (A.10)$$

The same procedure can be done in position-momentum space, too.

We get in the $q, p$ representation

$$\frac{\partial F^f(q,p)}{\partial t} = \hat{\Omega}_W F^f(q,p) + \left( \begin{array}{l} \omega \left( \begin{array}{l} 2\dfrac{\partial^2}{\partial p \partial q}(C_1 - C_2) + \\ 2iC_3\left(\dfrac{\partial^2}{\partial p^2} - \dfrac{\partial^2}{\partial q^2}\right) \end{array} \right) + \\ \gamma\left(-C_1\dfrac{\partial^2}{\partial p^2} - C_2\dfrac{\partial^2}{\partial q^2} + C_3 2i\dfrac{\partial^2}{\partial p \partial q}\right) + \\ 4A_x \left( \begin{array}{l} -\dfrac{\partial^2}{\partial p \partial q}(C_1 + C_2) + \\ i\left(\dfrac{\partial^2}{\partial p^2} + \dfrac{\partial^2}{\partial q^2}\right)C_3 \end{array} \right) + \\ 4A_y\left(-C_1\dfrac{\partial^2}{\partial p^2} + C_2\dfrac{\partial^2}{\partial q^2}\right) \end{array} \right) F^f(q,p) . \quad \text{(A.11)}$$

We see that in both representations, the propagator of the general Gaussian quasi-distribution function can be written as an exponent of a sum of operators representing the same Lie algebra we saw in the case of the Wigner function, and so again, the same technique can be applied in order to decompose it into simple propagators.